\documentclass[10pt,draft,reqno]{amsart}
     \makeatletter
     \def\section{\@startsection{section}{1}%
     \z@{.7\linespacing\@plus\linespacing}{.5\linespacing}%
     {\bfseries
     \centering
     }}
     \def\@secnumfont{\bfseries}
     \makeatother
\newtheorem{theorem}{Theorem}[section]
\newtheorem{lemma}[theorem]{Lemma}
\newtheorem{proposition}[theorem]{Proposition}

\theoremstyle{definition}
\newtheorem{definition}[theorem]{Definition}
\newtheorem{example}[theorem]{Example}
\theoremstyle{remark}

\numberwithin{equation}{section}
\usepackage{amssymb}
\usepackage{amsmath}
\usepackage{mathrsfs}
\usepackage{amsfonts}
\usepackage{dutchcal}
\usepackage{braket}
\usepackage{physics}
\usepackage{color}

\def\XXint#1#2#3{{\setbox0=\hbox{$#1{#2#3}{\int}$ }
\vcenter{\hbox{$#2#3$ }}\kern-.6125\wd0}}

\let\origmaketitle\maketitle
\def\maketitle{
  \begingroup
  \def\uppercasenonmath##1{} 
  \let\MakeUppercase\relax 
  \origmaketitle
  \endgroup
}

\begin {document}
\title{Supersymmetric Quantum Fields via Quantum Probability}

\author{Radhakrishnan Balu}
\address{Army Research Laboratory Adelphi, MD, 21005-5069, USA}
\email{radhakrishnan.balu.civ@army.mil}
\urladdr{http://sites.google.com/view/radbalu}

\subjclass[2010] {Primary Quantum theory; Secondary Relativity and gravitational theory}


\begin{abstract}
 The super version of imprimitivity theorem is available now to describe global supersymmetry of systems using the representations of super Lie groups (SLG). This result uses the equivalence between super Harish-Chandra pairs and super Lie groups at the categorigal level and is applicable to super Poincar\'e group and generalizes a smooth SI to super context. We apply the result to build supersymmetric quantum fields. Towards this end, we set up a super fock space of a disjoint union of super Hilbert spaces which is equivalent to super tensoring of boson (even) part symmetrically and that of fermion (odd) part antisymmetrically of the super particle Hilbert space. This leads to a super fock space that is disjoint union of bosonic and fermionic spaces, that is $Z_2$ graded. We derive covariant Weyl operators for light-like fields, with the massless super spinorial multiplet as an illustrative example. First, we build a representation of a light-like little group in terms of Weyl operators. We then use this construction to induce a representation of Poincar\'e group to construct the fields via super version of imprimitivity theorem.
\end{abstract}

\maketitle

\section{Introduction}
\label{intro}
Systems of imprimitivity (SI) is a way to characterize the unitary representations of a Lie group in a comprehensive way.  SI is a composite object $(G, \Omega)$ of a representation of a group $G$ and its action on a $G$-space $\Omega$ and we say it lives on $\Omega$. Mackey machinery is a set of techniques to induce representations of a group, from that of a subgroup $H$, that are systems of imprimitivity. The configuration spaces of interest to us in this work are orbits of little groups (space-like, time-like, and light-like) defined on the forward mass hyperboloid and the homogeneous space $G/H$ where $H$ is a closed subgroup of $G$ that consists of left cosets $gH, g \in G$. In the super context this homogeneous space stays the same as we assume the odd part of it as the entire super Lie algebra of $G$ that forms the habitat of super systems of imprimitivity (SSI) with different systems live on various orbits. When SI construction is applied to the Poincare group the projective unitary irreducible representations (PUIR) form the quantum states of fundamental particles with $\Omega$ being the configuration spacetime of the particles. From SI characterizations we can derive the canonical commutation relations and infinitesimal forms in terms of differential equations ($Schr\ddot{o}dinger$, Heisenberg, and Dirac etc) \cite {Wigner1949} \cite {Rad2019} and \cite {Rad2024}. The machinery originally applicable to semidirect products, Poincare group is a semidirect product of homogeneous Lorentz and $\mathbb{R}^4$, have been generalized in so many ways including a c*-algebraic version and a supersymmetric (SUSY) formulation. In this work, we will apply the SSI techniques to build Weyl operators on super fock spaces and then construct annihilation and creation field operators that are indexed by members of Poincare. In our earlier work we have constructed covariant Quantum Fields via Lorentz Group Representation of Weyl Operators \cite {Radbalu2020}. Here we generalize them to supersymmetry by building the super fock spaces for massless super multiplets (spinorial) and then the covariant filed operators. We glossed over domain consideration and we refer the reader to Varadarajan et.al \cite {Varadarajan2006} work for a detailed discussion that discuss the unbounded odd operators. Alternately, unbounded operators can be treated using affiliated operators that use bounded functions. More specifically, if $A$ is an unbounded operator we can consider $T_A = (A + i\mathbb{I})^{-1}$ that is bounded.

\begin {definition} A super Hilbert space is a $Z_2$-graded super vector space $\mathscr{H} = \mathscr{H}_1 \oplus \mathscr{H}_2$ over $\mathbb{C}$ with a scalar product $(. , .)$ where the $\mathscr{H}_i (i = 0, 1)$, referred as even and odd, are closed mutually orthogonal subspaces. We set up the parity operator as $$ p(x) = \begin{cases} 0, & \text{if x }\in \mathscr{H}_1, \\ 1, & \text{if x }\in \mathscr{H}_2. \end {cases}$$ We define an even super Hilbert form
$$\langle x,y\rangle = \begin{cases} 0, & \text{if x and y are of opposite parity} \\ (x,y), & \text{if x and y are even} \\ i(x,y), & \text{if x and y are odd} \end {cases}.$$
We have $$\langle y,x \rangle = (-1)^{p(x)p(y)}\overline{\langle x,y \rangle}.$$
\end {definition}
If $T(\mathscr{H} \rightarrow \mathscr{H})$ is a bounded linear operator, we denote by $T^*$ its Hilbert space adjoint and by $T^{\dag}$ its super adjoint given by $\langle Tx, y \rangle = (-1)^{p(T)p(x)}\langle x, T^{\dag} y \rangle$ .

\begin {definition} A super Lie group is $(G_0, \mathcal{g})$ is a a super Harish-Chandra pair if $G_0$ is a classical Lie group and $\mathcal{g}$ is a super Lie algebra with an action of $G_0$ on it such that
(i) Lie($G_0)$) = $\mathcal{g}_0$ = the even part of $\mathcal{g}$.
(ii) The action of $G_0$ on $\mathcal{g}$ is the adjoint action of $G_0$; more precisely, the adjoint action of $G_0$ on $\mathcal{g}$ is the differential of the action of $G_0$ on $\mathcal{g}$.
A representation of a super Lie group is a triple ($\pi, \gamma, \mathscr{H}$) where $\pi$ is an even representation of $G_0$ in a super Hilbert space $\mathscr{H}$ and $\gamma$ is a super representation of $\mathcal{g}$ in $\mathscr{H}.$
\end {definition}

\begin {definition} A super Lie algebra is a super vector space $\mathcal{g}$ with a bilinear bracket $[ , ]$ such that $\mathcal{g}_0$ is an ordinary Lie algebra with $[.,.]$ and $\mathcal{g}_1$ is a $\mathcal{g}_0$-module for the action $a \rightarrow ad(a): b \rightarrow [a, b], (b \in \mathcal{g}_1$). Further, $a \otimes b \rightarrow [a, b]$ is a symmetric $\mathcal{g}_0$-module map from $\mathcal{g}_1 \otimes \mathcal{g}_1$ to $\mathcal{g}_0$. It also satisfies the nonlinear condition $$[a, [a, a]] = 0, \forall g \in \mathcal{g}_1.$$ One way to ensure this last condition is met is to ensure that the range of the odd bracket $\mathcal{g}_2$ is a subset of $\mathcal{g}_0$ which acts on $\mathcal{g}_1$ trivially as $$\mathcal{g}_2 \subset \mathcal{g}_0 \Rightarrow [\mathcal{g}_1, \mathcal{g}_0] = 0.$$ A super algebra $A$ is an algebra of endomorphisms of linear maps on a super vector space $V$. The maps that preserve he grading of $V$ are designated as even and that reverse them are called odd. To get a super Lie algebra from $A$ we ca use the bracket $$[a, b] = ab - (-1)^{p(a)p(b)}ba$$. 

\end {definition}
Let us review the notions to describe systems of imprimitivity (SI) and an important result by Mackey that characterizes such systems in terms of induced representations, key notions in Clifford algebras, spinor fields, and Schwartz spaces \cite {Varadarajan1985} before discussing our main result in the super context. We provide the SUSY generalizations along with their classical counterparts using the notations and notions from the works of Varadarajan \cite {Varadarajan1985, Varadarajan2006, Varadarajan2011}.

\begin {definition} \cite {Varadarajan1985} A G-space of a Borel group G is a Borel space X with a Borel automorphism $\forall{g\in{G}},t_g:x\rightarrow{g.x},x\in{X}$ such that
\begin {align}
&t_e \text{ is an identity} \\
&t_{g_1,g_2} =t_{g_1}t_{g_2}
\end {align}
The group G acts on X transitively if $\forall{x,y\in{X}},\exists{g}\in{G}\ni{x=g.y}.$
\end {definition}
\begin {definition} \cite {Varadarajan1985} A system of imprimitivity for a group G acting on a Hilbert space $\mathscr{H}$ is a pair (U, P) where $P: E\rightarrow{P_E}$ is a projection valued measure defined on the Borel space X with projections defined on the Hilbert space and U is a representation of G satisfying
\begin {equation}
U_gP_EU^{-1}_g = P_{g.E}
\end {equation}  
\end {definition} 
Systems may be decomposed into SI $(G_0, \Omega = G_0/H_0)$ where $H_0$ closed subgroup of $G_0$ and a stabilizer at $\omega_0 \in \Omega$ on orbits by the transitive actions of the group and there exists a functor between the category of unitary representations of $H_0$ and the category of SI $(G_0, \Omega)$. In the case of Poincare group transitive SI is of interest to us we use the specialized version of the Mackey machinery. Let $\sigma$ be a representation of $H_0$ on a Hilbert space $\mathcal{K}^\sigma$ then there is a canonical SI $(\pi^\sigma, P^\sigma)$ for $G_0$ based on $\Omega$ with the representation induce by that of $H_0$ and the natural projection valued measure on $\mathcal{K}^\sigma$. The Hilbert space is the equivalent class of measurable functions $f: G_0 \rightarrow \mathcal{K}^\sigma$ satisfying:
\begin {align}
f(x\eta) &= \sigma(\eta)^{-1}f(x), \text{ for almost all }\eta \in H_0. \\
\int \abs{f(x)}^2_{\mathcal{K}^\sigma}dx & < \infty.
\end {align}
The representation $\pi^\sigma$ acts by left translation and the SI relation $\sigma \rightarrow (\pi^\sigma, \mathcal{K}^\sigma)$ states that there is functor exists between the category of  and the unitary representations of $H_0$ and the category of SI based on $\Omega$. One can develop an intuition \cite {Varadarajan2011} as $\mathcal{K}^\sigma$ as attached to the fixed point $\omega_0$ and for all the non-fixed points $\omega = g[\omega_0]$ a Hilbert space $\mathcal{K}^\sigma_\omega$ via an unitary isomorphism. This results in a fiber bundle $\mathcal{V}^\sigma = \mathcal{K}^\sigma \times G_0/ \sim$, where the equivalence relation is defined by $(g, \psi) \sim (g\eta, \sigma(\eta)^{-1}\psi)$. The group $G_0$ has a a natural right action on the bundle.

\begin {definition} \cite {Varadarajan2011} A super system of imprimitivity is a tuple $(\pi, \rho^\pi, \mathcal{H}, P)$ for a SLG $G = (G_0, \mathcal{g})$ living on $\Omega = G_0/H_0$ where  $H = (H_0, \mathcal{h})$ is a special subgroup of $G$ satisfying the following properties:\\
(1) The tuple $(\pi, \rho^\pi, \mathcal{H})$ is a unitary representation of the SLG $G$.\\
(2) The tuple $(\pi, \mathcal{H}, P)$ is a classical system of imprimitivity for $G_0$ in $\mathcal{H}$, based on $\Omega$, with $P$ an even operator.\\
(3) The projection valued measure $P$ commutes with $\rho^\pi$; that is,  the spectral projections of the odd algebra $\rho^\pi(X), X \in \mathcal{g}_1$ commute with the projections $P_E, E \subset \Omega$.
\end {definition}
The last condition may be unpacked by starting from the assumption that in the super context the configuration space $(\Omega = G/H = G_0 )/H_0$ is purely even. This implies $Xf = 0, X \in \mathcal{g}_1, f \in C_c^\infty(\Omega)$. Now, the commutation follows as $$[\rho^\pi (X), A(f)] = A(X(f)) = 0.$$

This definition of SSI lets us lift our earlier result on SI \cite {Radbalu2020} representation of classical Poincar\`e to the super context by retaining the even part and making sure that the odd part is compatible.

\section {Super fiber bundle representation and super semidirect products}

The states of a freely evolving relativistic quantum particles are described by unitary irreducible representations of Poincar\'e that has a geometric interpretation in terms of fiber bundles. 

\begin {definition}
\em {semidirect product of groups} Let A and H be two groups and for each $h\in{H}$ let $t_{h}:a\rightarrow{h[a]}$ be an automorphism (defined below) of the group A. Further, we assume that $h\rightarrow{t_h}$ is a homomorphism of H into the group of automorphisms of A so that
\begin {align} \label{semidirectEq}
h[a] &= hah^{-1}, \forall{a\in{A}}. \\
h &= e_H, \text{  the identity element of H}. \\
t_{{h_1}{h_2}} &= t_{h_1}t_{h_2}.
\end {align}
Now, $G=H\rtimes{A}$ is a group with the multiplication rule of $(h_1,a_1)(h_2,a_2) = (h_1{h_2},a_1{t_{h_1}}[a_2])$. The identity element is $(e_H,e_A)$ and the inverse is given by $(h,a)^{-1} = (h^{-1},h^{-1}[a^{-1}])$. 
\end {definition}
When $H$ is the homogeneous Lorentz group $L_0$ and A is the translation group $T = \mathbb{R}^4$ we get the Poincar\'e group $\mathscr{P}=H\rtimes{T}$ via this construction. The covering group of inhomogeneous Lorentz is also a semidirect product as $\mathscr{P}^* = H^*\rtimes \mathbb{R}^4$ and as every irreducible projective representation of $\mathscr{P}$ is uniquely induced from a representation of $\mathscr{P}^*$ we will work with the covering group, whose orbits in momentum space are smooth, in the following. We make the assumption that the semidirect product is regular in that for the action of $H$ on the dual $H^*$ there is a Borel cross section. In other words all the $H-$orbits in $H^*$ are locally closed \cite {Effros1965}.

We will need the following lemma for our discussions on constructing induced representations using characters of an abelian group as in the case of equation \eqref {eq: poissonBr}.
\begin {lemma} (Lemma 6.12 \cite {Varadarajan1985})
Let $h \in H$. Then, $\forall x \in \hat{A}$ where $\hat{A}$ is the set of characters of the group A (which in our case is $\mathbb{R}^4$), there exists one and only $y \in \hat{A}$ such that $y(a) = x(h^{-1}[a]), \forall a \in A$. If we write y = h[x], then $h,x \rightarrow h[x]$ is continuous from $H \times \hat{A}$ into $\hat{A}$ and $\hat{A}$ becomes a H-space. Here, y can be thought as the adjoint for action of H on $\hat{A}$ and the map $\hat{p}$ in equation\eqref {eq: poissonBr} is such an example that is of interest to our constructions. In essence, we have Fourier analysis when restricted to the abelian group A of the semidirect product.
\end {lemma}
\begin {definition}
A super translation group is a super Lie group $(T_0, \mathcal{t})$ where $T_0$ is abelian with a trivial action on the even part of the super Lie algebra $\mathcal{t}_0$. As $[\mathcal{t}_1, \mathcal{t}_1] \subset \mathcal{t}_0$ and $T_0$ acts trivially on $\mathcal{t}_1$ the nonlinear condition of super Lie algebra is satisfied. 
\end {definition}
Now, let us suppose that $\mathcal{t}_1$ is an $L_0$-module (Lorentz module) and that the super commutator map $a, b \rightarrow [a, b]$ is an $L_0$-invariant from $\mathcal{t}_1 \times \mathcal{t}_1$ into $\mathcal{t}_0$. Then, $\mathcal{g} = \mathcal{l}_0 \oplus \mathcal{t}$ is a super Lie algebra with $\mathcal{g}_0 = \mathcal{l}_0 \oplus \mathcal{t}_0 = Lie(G_0), \mathcal{t}_1 = \mathcal{g}_1$. Here again the odd bracket has $\mathcal{t}_0$ as range subset of $\mathcal{t}_0$ and acts trivially on $\mathcal{t}_1$ to make it as a super Lie algebra. This way we can construct a whole family of super Lie groups $(G_0, \mathcal{g})$ including super Poincare using homogeneous Lorentz and super translation group. we can generalize the SI example on $SO_3$ (or the double cover as the spin group) as the closed subgroup $S_0 \subset L_0$  and $H_0 = T_0S_0$ of $G_0$. SO, $(H_0, \mathcal{h}$ is a special super Lie group of $(G), \mathcal{g})$ with $\mathcal{h} = \mathcal{h}_0 \oplus \mathcal{t}_1$.

\section {Super Little groups (stabilizer subgroups)}
We consider different SI that live on the orbits of the stabilizer subgroups as concrete examples. It is helpful to have the picture that SI is an irreducible unitary representation of Poincar\'e group $\mathscr{P}^+$ induced from the representation of a subgroup of homogeneous Lorentz such as $SO_3$. Using the IRR $U_m, m \in SO_3$ we can induce a representation as $$U_{(m,g)}\psi(k) = e^{i\{k,g\}}\psi(R^{-1}_rk)$$ where g belongs to the $\mathscr{R}^4$ portion of the Poincar\'e group, m is a member of the rotation group, and the duality between the configuration space $\mathbb{R}^4$ and the momentum space $\mathbb{P}^4$ is expressed using the character the irreducible representation of the group $\mathbb{R}^4$ as:
\begin {align} \label {eq: poissonBr}
\{k,g\} &= k_0 g_0 - k_1 g_1 - k_2 g_2 - k_3 g_3, p \in \mathbb{P}^4. \\
\hat{p}:x &\rightarrow e^{i\{k,g\}}. \\
\{Lx, Lp \} &= \{ x, p \}. \\\
\hat{p}(L^{-1}x) &= \hat{Lp}(x).
\end {align}
In the above L is a matrix representation of Lorentz group acting on $\mathbb{R}^4$ as well as $\mathbb{P}^4$ and it is easy to see that $p \rightarrow Lp$ is the adjoint of L action on $\mathbb{P}^4$. The $\mathbb{R}^4$ space is called the configuration space and the dual $\mathbb{P}^4$ is the momentum space of a relativistic quantum particle with a real mass.

The stabilizer subgroup (Light-like particles) of the Poincar\'e group $\mathscr{P}^+$ at  $(1,0,0,1)$ \cite {Kim1991}: has generators in terms of the Pauli matrix $\sigma_3$, and two matrices $N_1 = \begin{bmatrix} 0 & 1 \\ 0 & 0 \end{bmatrix}, N_2 = \begin{bmatrix} 0 & i \\ 0 & 0 \end{bmatrix} $, that are a rotation around the Z axis and the boost $\Lambda_p$ in the spatial direction \cite {Kim1991}. There is a Lorentz frame where the momentum is proportional to $(1,0,0,1)$ but there is no rest frame for the particle. The stabilizer subgroup at this point is isomorphic to the two dimensional Euclidean group $E_2$, and the orbits of this group on $\mathbb{R}^4$ host SI. In the supersymmetric version only admissible orbits host SSI.

Let $L_0^\lambda, \lambda \in T_0^*$ be the stabilizer of $\lambda$ in $L_0$ and denote the corresponding super Lie algebra as 
\begin {equation} \label {eq: LieAlgebra}
\mathcal{g}^\lambda = \mathcal{t}_0 \oplus \mathcal{l}_0^\lambda \oplus \mathcal{g}_1, \mathcal{l}_0^\lambda = Lie(L_0^\lambda).
\end {equation}
Note that the odd part of the super Lie algebra is assumed to be that of the whole group Poincar\'e. Then, the little super group at $\lambda$ is $S^\lambda = (T_0 L_0^\lambda, \mathcal{g}^\lambda)$ which is a special super subgroup of $(G_0, \mathcal{g}).$ A unitary representation $(\sigma, \rho^\sigma)$ of super little group $S^\lambda$ is $\lambda$-admissible if $\sigma(t) = e^{i\lambda t}\mathbb{I}, \forall t \in T_0.$ The point $\lambda$ is admissible if there is an irreduicble unitary representation $S^\lambda$ that is $\lambda$-admissible. The set of all $\lambda$-admissible points that is also $L_0$-invariant ($L_0$-orbit) is denoted by $$T_0^+ = \{\lambda \in T_0^* | \lambda-admissible\}.$$ We can obtain a spectral measure $P$ on $T^*$ from an unitary representation $(\pi, \rho^\pi)$ of a super Lie group $S$ by restricting it to $T^*$ and Fourier transforming it as $$\pi(t) = \int_{T^*} e^{i\lambda t} dP(\lambda), t \in T^*.$$ 
\begin {theorem} \cite {Varadarajan2006} The spectrum of every irreducible representation of the super Lie group $S = (G_0, \mathcal{g})$ is in some orbit of $T^+_0$. For each orbit in $T^+_0$ and choice of $\lambda$ in it, the assignment that takes a $\lambda$-admissible unitary representation $\gamma = (\sigma, \rho^\sigma)$ of $S^\lambda$ into the unitary representation $U^\gamma$ of $(G_0, \mathcal{g})$ induced by it, is a functor which is an equivalence of categories between the category of the $\lambda$-admissible unitary representation of $S^\lambda$ and the category of unitary representation of $(G_0, \mathcal{g})$ with their spectra in their orbit. Varying $\lambda$ in that orbit changes into an equivalent one. In particular this functor gives a bijection between the respective sets of
equivalence classes of irreducible unitary representations.
\end {theorem}

One point worth noting is that if we restrict the UIR of $G_0, \mathcal{g}$ to $G_0$, that from supersymmetric Poincar\`e to regular Poincar\`e the URs will be on the same orbit. Moreover, the theorem says the restricted representations will in fact be UIRs. This implies that the super partners will have the same mass until SUSY is broken. 
\

\section {Super context: Clifford Algebras and Spinor Fields}
As we have to incorporate spinor filds in to the description of super particles whose symmetry is described by super Poincar\`e let us construct a representation for $\rho^\sigma$ with nondegenerate bilinear quadratic form. For the unitary representations $S^\lambda$ of interests to us satisfy $$\sigma(t) = e^{i\lambda(t)}\mathbb{I}, \lambda \in T^*$$ we can define the derivative $$-id(Z) = \lambda(Z)\mathbb{I}, (Z \in \mathcal{t}_0).$$ As $[X_1, X_2] \in \mathcal{t}_0) \forall X_i\in \mathcal{g}_1$ holds we have $$[\rho^\sigma(X_1), \rho^\sigma (X_2)] = 2\Phi_\lambda (X_1, X_2)\mathbb{I}, \forall X_i\in \mathcal{g}_1.$$
Now, $\Phi_\lambda$ is the symmetric bilinear form defined as $$\Phi_\lambda (X_1, X_2) = \frac{1}{2} \lambda([X_1, X_2])$$ on the algebra $\mathcal{C}_\lambda = \mathcal{g}_1 \otimes \mathcal{g}_1 / \sim, X^2 = Q_\lambda (X), Q_\lambda (X) = \Phi_\lambda(X, X), X \in \mathcal{g}_1.$
We can think of $\rho^\sigma$ as a representation of the Clifford algebra $\mathcal{C}_\lambda$. As $L_0$ acts on $\mathcal{g}_1$ with $Q_\lambda$ invariance we can extend the action $x, a \longmapsto x[a], x \in L_0, a \in \mathcal{g_1}$ to $\mathcal{C}_\lambda = \mathcal{g}_1 \otimes \mathcal{g}_1.$ We thus get a representation $S^\lambda = (\sigma, \tau)$ where $\tau$ is a self-adjoint representation of $\mathcal{g}_1$ and $\sigma$ a unitary representation of $T_0 L_0^\sigma$ that satisfies the relation $$\sigma(t) = e^{i\lambda(t)}\mathbb{I}, (t \in T_0), \tau(x[a]) = \sigma(a)\tau(x)\sigma(x)^{-1}, \forall a \in \mathcal{C}_\lambda, \forall x \in L_0^\sigma.$$
The $\lambda$-admissible IRR is equivalent to positive energy condition $Q_\lambda(X) \ge 0, X \in \mathcal{g}_1$ \cite {Varadarajan2011}, Hamiltonians are based on these operators, that exclude imaginary mass tachyons in SUSY context. In this work we are concerned with massless particles, SSI will live on zero mass orbits of the hyperboloid, to keep the set up simple.

Our requirement for super Poincar\`e group $(G_0, \mathcal{t})$ is that the even part of the super algebra $\mathcal{t}_0$ should have a non degenerate symmetric bilinear form and the odd part $\mathcal{t}_1$ should have a spin structure to incorporate fermionic statistics. In addition, $\mathcal{t} = \mathcal{t}_0 \oplus \mathcal{t}_1$ has to be a real $G_0$ module. Spin modules are not modules for orthogonal group (Lorentz) we have to consider its double cover $\mathcal{SL}(2, \mathbb{C})$ (the spin group). To have spinors the super Lie algebra part of the Harish-Chandra pair has to be real and we have to set $\mathcal{t}_1 = \mathbb{C}^2 \oplus \overline{\mathbb{C}}^2$, which would describe the majorana fermions, and complexify the group with two copies as $\mathcal{SL}(2, \mathbb{C})\otimes \mathcal{SL}(2, \mathbb{C})$. The number $N$ of spin modules determine the $N$-generators of the extended supersymmetry. The spin group embeds into the even part of the Clifford algebra we described earlier. This makes the Clifford algebra covariant with respect to Lorentz group which will enable the derivation of Dirac equation \cite {Varadarajan2011}. 

For our result that builds a representation for super Poincar\`e  on a super fock space we need a version of SSI theorem specific for this SLG that we state here without proof.

\begin {theorem} \cite {Varadarajan2006}
The irreducible unitary representations of a super Poincar\`e
group $S = (G_0, \mathcal{g})$ are parameterized by the orbits of $p$ with $p_0 \geq 0, \langle p, p \rangle \geq 0$, and for such $p$, by irreducible unitary representations of the stabilizer $L^p_0$
at $p$. Let $\tau_p$ be an irreducible SA representation of the Clifford algebra $\mathcal{C}_p$ and let $\kappa_p$ be the representation of $L^p_0$ in the space of $\tau_p$ defined earlier. Then,
for any irreducible unitary representation $\theta$ of $L^p_0$ the pair $(\sigma, \rho^\sigma)$ defined by
$$\sigma = e^{ip}\theta \otimes \kappa_p, \rho\sigma(x) = \mathbb{I} \otimes \tau_p, (x \in \mathcal{g}_1).$$
is an irreducible unitary representation of the little super group $$S^p = (T_0 L^p_0, \mathcal{l}_0 \oplus \mathcal{l}^p_0 \oplus \mathcal{g}_1)$$, and all irreducible unitary representations of $S^p$ are obtained in this manner. The unitary representation $\Theta_{\theta p}$ of the super Lie group $S$ induced by it is irreducible and all irreducible unitary representations of $S$ are obtained in this manner, the correspondence $(p, \theta) \longmapsto \Theta_{\theta p}$ being
bijective up to equivalence.
\end {theorem}
\
\section {Supersymmetric quantum fields}
A super Poincare SLG is a semidirect product between homogeneous Lorentz (classical group) and super spacetime translation group. That is, in SUSY for Poincar\`e the spacetime is augmented with fermionic degrees of freedom and the semdirect product with homogeneous Lorentz taken. We will stick to this setup but use a representation of Lorentz on a super fock space for the semidirect product. To keep things simple in terms of mass of super multiplets our super fock space will have particles of zero mass but with different spins.

Let us construct the single particle Hilbert space before second quantize the system. We need to describe few ingredients to construct the Hilbert space of a Weyl fermion (equivalent of fermionic photino) namely, the fiber bundle, the fiber vector space, an inner product for the fibers, and an invariant measure. 
The 3+1 spacetime Lorentz group $\hat{O}(3,1)$-orbits  of the momentum space $\mathscr{R}^4$, where the systems of imprimitivity established will live, described by the symmetry $\hat{O}(3,1)\rtimes\mathbb{R}^4$. The orbits have an invariant measure $\alpha^+_m$ whose existence is guaranteed as the groups and the stabilizer groups concerned are unimodular and in fact it is the Lorentz invariant measure $\frac{dp}{p_0}$ for the case of forward mass hyperboloid.  The orbits are defined as (we use the standard spin quantum number for the particles differing from Varadarajan who uses twice that number):
\begin {align}
X^{+,1/2}_m &= \{p: p^2_0 - p^2_1 - p^2_2 - p^2_3 = m^2, p_0 > 0\}, \text{\color{red} forward mass hyperboloid}.\\
X^{-,1/2}_m &= \{p: p^2_0 - p^2_1  - p^2_2 - p^2_3 = m^2, p_0 < 0\}, \text{\color{red} backward mass hyperboloid}. \\
X_{00} &= \{0\}, \text{\color{red} origin}.
\end {align}
Each of these orbits are invariant with respect to $\hat{O}(3,1)$ and let us consider the stabilizer subgroup of the first orbit at p=(1,1,0,0). Now, assuming that the spin of the particle is 1 and mass $m \rightarrow 0$ (massless fermion) let us define the corresponding fiber bundles (vector) for the positive mass hyperboloid that corresponds to the positive-energy states by building the total space as a product of the orbits and the group $SL(2, C)$.
\begin {align}
\hat{B}^{+,1/2}_m &= \{(p,v) \text{   }p\in{\hat{X}^{+1}_m, }\text{   }v\in\mathscr{C}^4,\sum_{r = 0}^3 p_r \gamma_r v = 0\}. \\
\hat{\pi} &: (p,v) \rightarrow {p}. \text{  Projection from the total space }\hat{B}^{+,1/2}_m \text{ to the base }\hat{X}^{+,1/2}_m.
\end {align}
In case of a spin $1$ particle \cite {Varadarajan1985} we will have to construct the bundle as follows, symmetric tensor of two half spins, and we will make a remark about it later in the context of second quantization:
\begin {align} \label{eq: higher spins}
\hat{B}^{+,1}_m &= \{(p,v) \text{   }p\in{\hat{X}^{+1}_m, }\text{   }v\in\mathscr{C}^4 \otimes \mathscr{C}^4,\sum_{r = 0}^3 p_r \gamma_r v = 0, \sum_{r = 0}^3 p_r \gamma_r^2 v = 0\}. \\
\hat{\pi} &: (p,v) \rightarrow {p}. \text{  Projection from the total space }\hat{B}^{+,1/2}_m \text{ to the base }\hat{X}^{+,1/2}_m.
\end {align}
It is easy to see that if $(p, v) \in B_0^{+,1/2}$ then so is also $(\delta(h)p, S(h^{*-1})v)$. Thus, we have the following Poincar\'e group symmetric action on the bundle that encodes spinors into the fibers:
\begin {equation}
h,(p,v) \rightarrow (p,v)^h = (\delta(h)p, S(h^{*-1})v).
\end {equation}
For $m > 0$ the fiber of $B_M^{+,1/2}$ at $p^(m) = ((a + m^2)^{1/2}, 1, 0, 0)$ is spanned by the vectors
\begin {align}
v_1^{(m)} &= \frac{1}{2}me_1 + \frac{1}{2}(1 + (1 + m^2)^{1/2})e_3. \\
v_2^{(m)} &= \frac{1}{2}me_4 + \frac{1}{2}(1 + (1 + m^2)^{1/2})e_2.
\end {align}
When we take the limit $m \rightarrow 0+$ $v_1,V-2$ converge to $e_3,e_2$ that space the fiber of $B_0^+$ at (1,1,0,0). The covering group $H^*$ is transitive on $X_m^+, X_0^+$ implies that the same convergence is true for any point. That is, if $p \in X_0^+$ then there are points $p^{(m)} \in X_m^+$ that converge to p as $m \rightarrow 0+$. This has the property that any vector $v$ in the fiber of $B_0^+$ at p can be expressed as the limit of $v^{(m)}$ which is in fiber of $B_m^{1/1}$ at $p^{(m)}$. The same set of arguments can be applied to $B_{-m}^{+,1/2}$ implying that the bundles $B_{-m}^{+,1/2}$ also converge to $B_0^+$.
The endomorphism (chirality or helicity operator) $\Gamma = i\gamma_0\gamma_1\gamma_2\gamma_3$ transforms $B_m^{+,1/2}(p) \rightarrow B_{-m}^{+,1}(p), \forall p \in X_m^+, m > 0$ as it anticommutes with all the gammas. In the limit $\Gamma$ leaves the fibers of $B_0^+$ invariant leading to higher degenerecies with $\Gamma = \begin{bmatrix} 1 & 0 \\ 0 & -1 \end{bmatrix}$. This means $\Gamma$ commutes with all of $S(h)$ implying that $(p, v) \in B_0^+ \Rightarrow (p, \Gamma v) \in B_0^+$. If we impose either of the condition $\Gamma \psi = \pm \psi$ then we can use 2x2 the Pauli matrices for the $\gamma$s and we get the description for a Weyl fermion.

$\Gamma$ has eigen values $\pm 1$ at fiber (1, 0, 0, 1) and hence is true of all fibers. The stability group $E^*$ at (1, 0, 0, 1) given by the matrices $\begin{bmatrix} z & a \\ 0 & (z^{-1}) \end{bmatrix}, z,a \in \mathscr{C}, \abs{z}=1$, induces a representation on the fibers as $m_{z,a} \rightarrow z^{\pm 1}$.
Next step is to ensure that there is an Hermitian form on the fibers that is positive definite and left invariant with respect to S. The form $v\rightarrow p_0^{-1}\langle v, v \rangle$ can be shown to satisfy the condition and by letting $m \rightarrow 0+$ the form is still invariant and we have $E_2$ is the stabiliser group at (1,0,0,1).

Now, we can define the states of the light like particles on the Hilbert space $\hat{\mathscr{H}}^{+,\mp 1/2}_0$, square integrable functions on Borel sections of the bundle $\hat{B}^{+,1}_0 = \{(p,v) : (p,v) \in B_0^+, \Gamma v = \mp v\}$.

The states of the particles are defined on the Hilbert space $\hat{\mathscr{H}}^{+,1/2}_m$, square integrable functions on Borel sections of the bundle $\hat{B}^{+,2}_m$ with respect to the invariant measure $\beta^{+,1}_0$, whose norm induced by the inner product is given below:
\begin {equation} \label {eq: section}
 \norm{\phi}^2 = \int_{X^+_m}p_0^{-1}\langle\phi{p},\phi{p}\rangle.{d\beta}^{+,1}_0(p).
\end {equation}
The invariant measure and the induced representation of the Poincar\'e group from that of the Weyl fermion are given below:
\begin {align}
{d\beta}^{+,1}_0(p) &= \frac {dp_1 dp_2 dp_3} {2(p_1^2 + p_2^2 + p_3^2)}. \\
(U_{h,x}\phi)(p) &= \text{exp i}\{x, p\} \phi (\delta(h)^{-1}p)^h.
\end {align}

\
\begin {definition} A super fock space $\Gamma$ of a super Hilbert space $\mathcal{H}$ is a disjoint union of bosonic fock space of $\Gamma_s(\mathcal{H})$ and fermionic fock space $\Gamma_a(\mathcal{H})$. That is, the even part of the Hilbert space are tensored symmetrically and the odd part of the Hilbert space are tensored antisymmetrically and a disjoint union is formed. The even part of the super fock space supports Weyl operators just as in the classical case. The odd operators of the super Hilbert space are tensored to form the odd operators of the SUSY system. 
\end {definition}
Let us now state and discuss the main result for the case of massless super multiplets with light-like momentum by constructing a strict cocycle from the representation of a subgroup following the prescription (lemma 5.24) in Varadarajan{'}s text. The SI is a consequence of strict cocycle property and  the construction is not canonical.

Lemma 5.24 \cite{Varadarajan1985}: 
Let $m$ be a Borel homomorphism of $H_0$, which is a subgroup of $G$ into $M$. Then there exists a Borel map $b$ of $G$ into $M$ such that 
\begin {align} \label {eq: cocycle} 
b(e) &= 1. \\
b(gh) &= b(g)m(h), \forall (g,h) \in G \times G_0.  
\end {align}
Corresponding to any such map b, there is a unique strict (G, X)-cocycle
f such that
\begin {equation} \label {eq: cocycle2}
f(g, g^1) = b(gg^1)b(g^1)^{-1}.
\end {equation}
$\forall (g,g^{-1}) \in G \times G$. f defines m at $x_0$. Conversely, when f is a strict (G, X)-cocycle and b is a Borel map such that it satisfies equation \ref {eq: cocycle} pair, then the restriction of b to $H_0$ coincides with the homomorphism m defined of at $x_0$ and b satisfies equation \ref {eq: cocycle2}.

\begin {theorem} Light-like Weyl representation $(G_0, \mathcal{g})$ of super Poincar\`e group on the super fock space $\Gamma = \Gamma_s(\hat{\mathscr{H}}^{+,m,1/2}_0) \oplus \Gamma_a(\hat{\mathscr{H}}^{+,m,1/2}_0)$ is a transitive super system of imprimitivity $(\pi, \rho^\pi, \Gamma, P)$ that lives on $\Omega = G_0/H_0)$. This is a system of N-extended SUSY with a super multiplet of single boson and a fermion.
\end {theorem}

\begin {proof}
We set up the super fock space as a disjoint union of $\Gamma_s(\hat{\mathscr{H}}^{+,m,1/2}_0)$ and $\Gamma_a(\hat{\mathscr{H}}^{+,m,1/2}_0)$ and the odd operators (super charges of SUSY) are defined between them (specifically, between the spacetime components of the sections of the bundles):
\begin {align*}
\Gamma_s(\hat{\mathscr{H}}^{+,1}_0) &= \oplus_{n = 0}^\infty (\hat{\mathscr{H}}^{+,m,1/2}_0)^{\otimes n}. \\
(\hat{\mathscr{H}}^{+,m,1/2}_0)^{\otimes n} &= \{ u \in (\hat{\mathscr{H}}^{+,m,1/2}_0)^{\otimes n}, U_\sigma u = u, \forall \sigma \in S_n\} \\
&\text{  the permutation group}.\\
U_\sigma u_1\otimes\dots\otimes u_n &= u_{\sigma^{-1}(1)}\otimes\dots\otimes u_{\sigma^{-1}(n)}.
\end {align*}
\begin {align*}
\Gamma_a(\hat{\mathscr{H}}^{+,m,1/2}_0) &= \oplus_{n = 0}^\infty (\hat{\mathscr{H}}^{+,m,1/2}_0)^{(a)\otimes n}. \\
(\hat{\mathscr{H}}^{+,m,1/2}_0)^{(a)\otimes n} &= \{ u \in (\hat{\mathscr{H}}^{+,m,1/2}_0)^{\otimes n}, U_\sigma u = \epsilon(\sigma) u, \forall \sigma \in S_n\} \\
&\text{  the permutation group}.\\
\epsilon(\sigma)   &= \pm 1 \\
&\text {  based on whether the permutation is even or odd}.
\end {align*}
In the above construction we used the same single particle Hilbert space (same spinors) to construct the bosonic and fermionic fock spaces respectively. Without second quantization, the single particle Hilbert space for bosons and fermions would differ in the dimension of the spinors used in the fibers as the higher spin partner of the multiplet will have a symmetric tensor product of spin components (equation \eqref{eq: higher spins}). We avoid that step by by symmetric fock space construction of the entire Hilbert space. Our construction facilitates applying boson-fermion correspondence  to build supercharges as in the second example below.

Let us first construct the classical SI $(\pi, \Gamma, P)$ on the super fock space with the homomorphism $g:L^p_0 \otimes L^p_0 \rightarrow U_g(\hat{\mathscr{H}}^{+,1}_0)$ from the two dimensional Euclidean group $L^p_0 = E_2$ to the unitary representation of the group in $\hat{\mathscr{H}}^{+,1}_0$. 
We note that it is a stabilizer subgroup which is also closed at the momentum $p = (m,0,0,m)$ and so $H/L^p_0$ is a transitive space and so the super version is a special subgroup (the odd part is the whole super Lie algevra).

Consider a map, from the light-like particle Hilbert space, $v(g):L^p_0 \rightarrow \hat{\mathscr{H}}^{+,1/2}_0$ satisfying the first order cocycle relation $v(gh) = v(g) + U_g v(h), g,h \in L^p_0$.
An example of such a map is the following: \cite {KP1992} 
\begin {align*}
\mathscr{H} &= \oplus_{j=0}^\infty \mathscr{H}_j. \\
H &= 1 \oplus \oplus_{j=1}^\infty H_j. \\
U_t &= e^{-itH}, t \in L^p_0. \\
v(t) &= tu_0 \oplus \oplus_{j=1}^\infty (e^{-itH_j} u_j
- u_j).
\end {align*}

Now, we can define the Weyl operator $V_g = W_g (v(g), U_g)$ where $g \in L^p_0$ for even part of the super fock space $\Gamma_s(\hat{\mathscr{H}}^{+,1/2}_0)$.

This is a projective unitary representation satisfying the commutator relation $V_g V_h = e^{iIm\langle v_g, U_g v_h \rangle} V_h V_g$ and let us denote the homomorphism from $L^p_0$ to $V_g$ by m. This guarantees a map (lemma 5.24, \cite {Varadarajan1985}) b that satisfies $b(gh) = b(g)m(h), g \in G_0, h \in L^p_0$ and such map can be constructed by considering the map
$c(x \rightarrow c(x)$ as Borel section of $\mathscr{P} / L^p_0$ (the choice of this section not a canonical one but immaterial to our purpose here) with $c(x_0) = e$. The map
$\beta$ maps $g \in G_0 \rightarrow g L^p_0$
\begin {align}
a(g) &= c(\beta (g))^{-1}. \\
b(g) &= m(a(g)). 
\end {align}
Then the strict cocycle $\phi$ satisfies $\phi(g_1, g_2) = b(g_1 g_2)b(g_2)^{-1}$.

We can now set the SI relation using the above cocycle as follows:
\begin {align*}
U_{h, x}\phi(p) &= e^{i\{x, p\}} \phi(g, g^{-1}x) f(g^{-1}x),  f\in\mathscr{H} \\
&\text {  character representation is defined in equation } \eqref{eq: poissonBr}.\\
P_E F &= \chi_E f \text { Position  operator}. \\
\end {align*}

We can construct the conjugate pair of field operators for the Fock space $\Gamma_s(\hat{\mathscr{H}}^{+,1/2}_0)$ as follows:

Let $p_g$ be the stone generator for the family of operators $P_{gt,p}, g \in \mathscr{P}, t \in \mathbb{R}$ and q(g) = p(ig) and we get the creation and annihilation operators as $a(g)^\dag = \frac {1}{2} ( q(g) - i p(g)$ and  $a(g) = \frac {1}{2} ( q(g) + i p(g)$.

With the identification $\pi$ as the Weyl representation on the even part of the super fock space we have the tuple $(\pi, \Gamma, P)$.
We can set up the representation $\rho^\pi$ on the super fock space of the super Lie algebra as per the equation \eqref {eq: LieAlgebra}. The third condition on SSI is also satisfied with our selection of purely even super homogeneous subgroup.

We can lift the representation to $\pi$ to $L_0^p \otimes L_0^p$ and more precisely to its double cover $SL(2, \mathbb{C} \otimes SL(2, \mathbb{C})$. Now, the super Lie algebra $\mathcal{t}_1$ of odd operators of the SLG will be a module of this spin group and we get the spinors that form the fibers of the bundle whose base is the orbit $\Omega$.
$\blacksquare$
\end {proof}

\begin {example}
SUSY systems can be set up by defining odd operators of the super Lie algebra. In the case of super Poincar\'e group  the odd and even sectors of the Hilbert space are based on spacetime translations group as the Lorentz group part of the semidirect product is the same in both cases. So, if we define the odd operators between Hamiltonians, that are based on spacetime coordinates, of the two sectors we get SUSY systems.
As an example, we can define a SUSY system of $N = 1$ with the configuration space $\mathbb{R}^4$ of the super particles with the Hamiltonian (summation convention assumed)
\begin {align*}
H &= \frac {1}{2} \{Q, Q^\dag \}. \\
Q &= (p_i - i\partial_i h(\psi^i)), i = 1, ... 4.\\
Q^\dag &= (p_i + i\partial_i h(\psi^{i\dag})), i = 1, ... 4.\\
& \text{In the above h is a real function and } \psi^i \in \Gamma_s(\hat{\mathscr{H}}^{+,m,1/2}_0), \psi^{i\dag} \in \Gamma_a(\hat{\mathscr{H}}^{+,m,1/2}_0).
\end {align*}
It is straight forward to generalize the above SUSY to an arbitrary manifold, with an abelian group structure, from $\mathbb{R}^4$ by defining p-forms (when p is even it is a boson and fermion otherwise) and the exterior derivatives used to define the supercharges. \cite {Witten1982}. In the above example, 0-forms are used to construct the supercharge $Q_1$ that maps bosons to fermions and $Q_2$ is its conjugate.

More general super multiplets can be defined with mappings between the real Clifford algebra modules that form the fibers of the boson (symmetric tensor) and fermion (antisymmetric tensor) bundles.
\end {example}
\begin {example}
The unification of the \cite {Hudson1986} two bosonic and fermionic processes by the parity process that $\mathbb{Z}_2$-grades the fock space up to time $t$ as, $dB= (-1)^\Lambda dA$, where $dB, dA, \Lambda4$ are bosonic, fermionic, and number processes respectively, inspires building SUSY systems at the second quantized operators level as we detail now.
We first rewrite the above equations as $$dB= \frac{1}{2}(1 \pm  (-1)^{\Lambda_H}) dA, H = a^\dag a$$ and using bosom field operators we can set up an extended $N = 2$ SUSY system with the following pair of supercharges that commute with a pair of Hamiltonians:
\begin {align*}
Q_{\pm} &= a \frac{1}{2}(\mathbb{I} \pm 2\pi i \Lambda), \Lambda = a^\dag a. \\
Q^{\dag}_{\pm} &= a^\dag \frac{1}{2}(\mathbb{I} \pm 2\pi i \Lambda). \\
H_{\pm} &= \frac{1}{2} \{a^\dag + a\} + \frac{1}{2}\pm 2\pi i\Lambda [a, a^\dag]. \\
Q_{\pm}^2 & = 0. \\
(Q^\dag_{\pm})^2 & = 0.\\
\{ Q_{\pm}, Q^\dag_{\pm} \} &= H_\pm.\\
[Q_{\pm}, H_{\pm}] &= 0. \\
[Q^\dag_{\pm}, H_{\pm}] &= 0. 
\end {align*}
\end {example}
Let us construct the same supercharges using quantum stochastic calculus by suitably modifying the example 25.18 in Parthasarathy{'}s book \cite{KP1992}.
Our starting point is the quantum stochastic differential equation involving the conservation process $\Lambda_1$ and the reflection process defined in terms of second a quantized unitary $J(t) = \Gamma(U_t), U_t = -1$ as:
\begin {equation}
dJ = -2J d\Lambda_1.
\end {equation}
The modified maps from bosonic creation and annihilation processes to fermionic process can be set up as:
\begin {align} \label{supercharge}
F_m (t) &= \int_0^t \frac{1}{2} (\mathbb{I}+ J) dA_m. \\
F^\dag_m (t) &= \int_0^t \frac{1}{2} (\mathbb{I}+ J)dA^\dag_m.
\end {align}
The next step is to establish proposition and we outline the steps here.
\begin {proposition} (25.19 \cite {KP1992})
$$\langle J(t)e(u), F_m(t)e(v)\rangle + \langle F^\dag_m (t)e(u), J(t)e(v)\rangle = 0.$$
\end {proposition}
We apply the proof outlined in \cite {KP1992} to our set up:
\begin {align}
\phi(t) &= \langle J(t)e(u), F_m(t)e(v)\rangle. \\
\psi(t) &= \langle F^\dag_m(t)e(v)\rangle, J(t)e(u)\rangle.
\end {align}
By applying Ito{'} rule, the unitarity of $J(t)$, and equations $(25.25)$ and $(25.26)$ in \cite {KP1992} we get:

\begin {align}
\phi(t) &= \frac{1}{2}\langle\langle u, v\rangle\rangle + \frac{1}{2}\langle\langle u, v\rangle\rangle - \int_0^t \phi(s) \langle\langle u, v \rangle\rangle(ds). \\
\psi(t) &= \frac{1}{2}\langle\langle u, v\rangle\rangle - \frac{3}{2}\langle\langle u, v\rangle\rangle - \int_0^t \psi(s) \langle\langle u, v \rangle\rangle(ds).
\end {align}
Adding the above two equations and using the fact that $\langle\langle u, v \rangle\rangle$ is non-atomic, we get
\begin {align}
\phi(t) + \psi(t) &= \{\phi(0) + \psi(0)\} exp\{-1\langle\langle u, v\rangle\rangle[0, t] \} = 0.
\end {align}
In the above, $\xi : \mathscr{F}_{\mathbb{R}_+} \rightarrow \mathscr{P}(\mathscr{H})$ is a fixed observable with no jumps and $m_t$ is a $\xi$-martingale defined as $$\xi ([0, s])m_t = m_s, \forall s < t.$$. The measure $\langle\langle u, v \rangle\rangle$ is defined as $$\langle\langle u, v\rangle\rangle[0, t] = \langle u_t, v_t\rangle\, \forall t > 0, u_{t]} = \xi ([0, t])u, t \ge 0$$.

By changing $(I + J)$ to $(I - J)$ in equations \eqref{supercharge} we get another supercharge for the $N = 2$ SUSY system.
\

\section {Summary and conclusions}
We derived the covariant super field operators in a Minkowsky space using induced representations of groups and  expressed them in terms of super systems of imprimitivity. We established the results for a simple   super multiplet by inducing a representation of Poincar\'e group from the subgroup that is a stabilizer at the momentum (m,0,0,m).This sets the stage for studying SUSY breaking in Minkowskian signature using the tools of SSI.

\par\bigskip\noindent
{\bf Acknowledgment.} The author acknowledges the suggestion from Michael Montgomery that affiliated operators could be used to deal with odd operators of the super Lie algebras that are unbounded. The author is grateful to Prof. Accardi who noted, at the Oaxaca workshop, the need to have the same single particle Hilbert space for both the fermionic and bosonic fock spaces in order to apply the boson-fermion correspondence upon second quantization.

\bibliographystyle{amsplain}

\end{document}